\documentclass[11pt,a4paper]{article}

\usepackage[T1]{fontenc}
\usepackage[utf8]{inputenc}
\usepackage{lmodern}
\usepackage[english]{babel}

\usepackage[margin=2.5cm]{geometry}
\usepackage{amsmath,amssymb}
\usepackage{graphicx}
\usepackage[font=small,labelfont=bf]{caption}
\usepackage{booktabs}
\usepackage{microtype}
\usepackage[numbers,sort&compress]{natbib}
\usepackage[colorlinks=true,allcolors=blue]{hyperref}

\graphicspath{{figures/}}

\title{\bfseries 3D-printed experimental devices for teaching\\ dynamical systems in physics}

\author{%
	Mario Silva\,$^{1}$\thanks{Corresponding author: \texttt{msilva@docente.ceibal.edu.uy}}
  \quad
  Victoria L\'opez\,$^{1}$
  \quad
  Daniel Freire\,$^{2}$
  \quad
  Arturo C. Marti\,$^{2}$
}

\date{%
  $^{1}$\,\textit{Instituto de Formaci\'on Docente de Salto, Consejo de Formaci\'on en Educaci\'on, ANEP, Salto, Uruguay}\\[2pt]
  $^{2}$\,\textit{Facultad de Ciencias, Universidad de la Rep\'ublica, Montevideo, Uruguay}\\[8pt]
  \today
}

\begin{document}

\maketitle

\begin{abstract}
\noindent
We present the design, construction and classroom use of 3D-printed experimental devices
aimed at introducing dynamical systems in physics courses. Theoretical, technical and
pedagogical criteria are proposed for producing low-cost, modifiable resources suitable for
experimental work. Three devices are analysed as examples: a magnetic pendulum, a wooden
woodpecker toy and a double pendulum. The suggested activities are organised around
prediction, observation, comparison between configurations, and the elaboration of physical
explanations. The main contribution of this work is to show concrete implementation
possibilities in secondary education, pre-service teacher training and introductory
university courses, together with a discussion of the scope and limitations of 3D printing
in experimental teaching.

\vspace{0.8em}
\end{abstract}

\section{Introduction}

Experimental work occupies a central place in physics teaching, because it links concepts,
models and observations in concrete situations. In the classroom, experimental devices
encourage students to describe phenomena, formulate predictions, compare behaviours and
revise explanations on the basis of evidence. This dimension becomes particularly relevant
for content in which the mathematical formalism may appear before students have built a
physical understanding of the phenomenon, as happens with systems exhibiting regularities,
transitions, or behaviours that are hard to anticipate by direct inspection.

A productive way of opening these discussions is to start from objects that students already
know from everyday life, and to turn them into quantitative experimental situations. Several
recent proposals aimed at the same educational levels addressed here illustrate this
strategy. The dynamics of a yo-yo, for instance, can be characterised in detail using the
gyroscope sensor built into a smartphone, which turns a familiar toy into a system where
rotational quantities become measurable~\citep{salinas2020analyzing}. An ordinary bottle of
tea, as it is progressively emptied, behaves as a Helmholtz resonator whose frequency shifts
in a way students can hear and then model~\citep{monteiro2018bottle}. The flight of a
commercial drone provides an accessible entry point to discuss thrust, torque balance and
stability~\citep{monteiro2022simple}, and the cooling of an object under forced convection
can be studied with equipment as simple as a fan and a temperature
sensor~\citep{dutra2024cooling}. In all of these cases the motivating power of the everyday
object is used deliberately: curiosity about a familiar artefact is redirected towards the
identification of variables, the formulation of models and the comparison between prediction
and measurement. The devices presented in this work follow the same rationale, extending it
to phenomena in which nonlinearity plays an essential role.

In many educational contexts, access to specific experimental equipment remains a
difficulty. Teaching laboratories often have general-purpose material for experimental work,
but they do not always have devices designed to explore phenomena that are less common in
introductory courses, such as nonlinear dynamics, sensitivity to initial conditions, or
limited predictability.

Faced with this situation, 3D printing offers a concrete alternative for designing, adapting
and building low-cost experimental resources. Several works have pointed out that this
technology makes it possible to produce customised parts, repair components, modify
dimensions and share digital models among teachers and
institutions~\citep{coakley2016printing,kolitsky2014reshaping,larnder2020physics}.

Devices inspired by \emph{physics toys}---that is, toys used for teaching purposes---are
valuable resources for introducing physical phenomena through objects that are manipulable,
visually appealing and easy to build. Some physics toys display conceptually rich behaviour,
because they involve friction, impacts, oscillations, changes of stability, or responses that
are sensitive to small variations in the initial
conditions~\citep{aravind2015inexpensive,leine2003nonlinear}. These features make them
suitable mediators for discussing ideas connected with dynamical systems, linearity,
nonlinearity and deterministic chaos~\citep{strogatz2018nonlinear,parlati2024design}.

The use of such devices can be integrated into active learning methodologies, in which
students take part in the construction of explanations rather than merely watching a
demonstration. Research in physics education has shown the value of interactive-engagement
strategies for promoting conceptual understanding~\citep{hake1998interactive,redish2003teaching}.
Within that framework, the Predict--Observe--Explain (POE) approach is a particularly
relevant option: before interacting with the device, students anticipate its behaviour;
during the experience, they contrast their predictions with what they observe; afterwards,
they elaborate explanations and revise their initial models~\citep{white2014probing}. Other
strategies may also be combined, such as inquiry activities, analysis of experimental
records, peer discussion, or guided modelling.

This article presents a proposal for designing, building and using 3D-printed experimental
devices aimed at addressing dynamical systems in physics courses. Our goal is to show how
this technology can favour the production of accessible, modifiable materials suitable for
promoting conceptual discussion in the classroom. To that end, we establish working criteria,
analyse three devices we have developed, and propose guidelines for their implementation,
considering both their scope and their limitations.

\section{3D printing and the design of experimental devices}

Incorporating 3D printing into physics teaching broadens the possibilities for building
resources for experimental work. In the laboratory, this technology has been identified as an
alternative for producing customised equipment, reducing costs and manufacturing parts
adjusted to specific needs~\citep{coakley2016printing}. In educational contexts, it allows a
shift from using finished objects to producing materials conceived for a teaching
purpose~\citep{kolitsky2014reshaping}. Moreover, the perspective of open, printable
laboratories reinforces the possibility of sharing, modifying and reproducing resources
across institutions, provided that the files are accompanied by assembly and usage
guidelines~\citep{mcnair2024toward}.

In this proposal, 3D printing is understood as a tool for building experimental devices
rather than as an end in itself. The main interest does not lie in the printed part, but in
the situation that this part makes possible in the classroom. For this reason, the design
should start from teaching questions: which phenomenon do we want to analyse? which variables
do we want to modify? which aspects of the system's behaviour do we want to make visible?
and what kind of student activity do we want to promote?

Developing these devices requires articulating conceptual, technical and classroom criteria.
Conceptually, the resource must allow a clear identification of the phenomenon under study
and offer conditions for repeating observations, comparing behaviours or modifying
parameters. Technically, it must be printable with accessible materials, require simple
assembly, and present a structure robust enough for repeated use. From the teaching point of
view, it must favour the formulation of questions, the recording of observations and the
construction of explanations. These criteria prevent 3D printing from being reduced to the
production of attractive but experimentally shallow objects.

A printed object acquires value as a teaching resource when it enables a situation in which
it is possible to observe a phenomenon, intervene on some of its conditions, and discuss
explanatory models. A change in the length of a part, in the position of a support, or in the
inclination of a surface can transform the observable behaviour of the system. This
possibility of modifying conditions turns the device into a tool for exploring physical
relationships, instead of presenting it as a demonstration.

The development of these resources can be organised as an iterative process. A first version
allows one to check the feasibility of the mechanism and to detect assembly problems. Later
versions can improve durability, reduce material consumption, simplify assembly, or increase
the visibility of the phenomenon for the whole group of students. The digital nature of the
models also makes it easier for other teachers to modify the files, print new versions and
adapt the device to their own classroom
conditions~\citep{coakley2016printing,mcnair2024toward}.

For devices aimed at the study of dynamical systems, these criteria become especially
important. Small variations in the initial conditions, in the geometry, or in the assembly
parameters can modify the observed response. This feature offers a concrete route for
discussing predictability, stability and nonlinearity using systems that are simple to build
yet conceptually rich~\citep{strogatz2018nonlinear,parlati2024design}.

\section{The 3D-printed devices}

The devices were selected on the basis of their potential for introducing dynamical phenomena
through manipulable objects that are easy to build and whose non-trivial behaviour is
apparent to the naked eye. The use of physics toys in teaching has a well-established
tradition that highlights their value for arousing curiosity, prompting observations and
opening conceptual discussions from simple
objects~\citep{aravind2015inexpensive,schlichting1987chaotic}. In this work, the design
sought to ensure that each device allows the phenomenon to be observed directly, without
initially requiring advanced mathematical treatment. This decision helps the experimental
experience function as a starting point for formulating questions, comparing behaviours and
constructing physical explanations.

The three devices presented here---a magnetic pendulum, a woodpecker toy and a double
pendulum---share several design features. All of them combine 3D-printed parts with readily
available materials such as strings, small masses, magnets, springs, rods and bearings. In
addition, they allow relevant system parameters to be modified, among them the initial
position, the length of the arms, the stiffness of the spring, the inclination of the pole,
or the distribution of the magnets. These variations make it possible to observe changes in
the response of the system and to discuss ideas associated with stability, predictability and
nonlinearity~\citep{strogatz2018nonlinear,parlati2024design}.

\subsection{Magnetic pendulum}

\begin{figure}[htbp]
  \centering
  \includegraphics[width=0.45\textwidth]{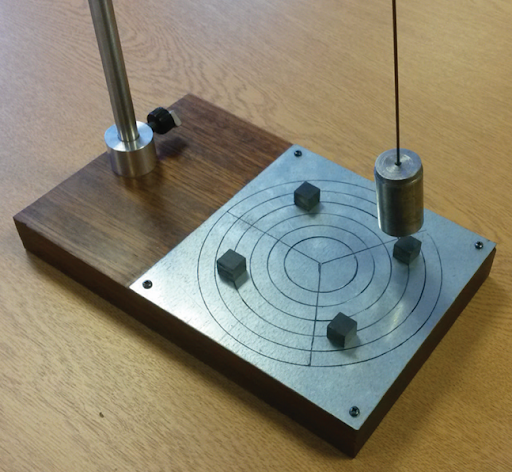}
  \caption{Setup of a magnetic pendulum with four magnets on the base.
           Reproduced from Ref.~\citep{christian2020chaos}.}
  \label{fig:magpend-scheme}
\end{figure}

The magnetic pendulum consists of a suspended ferromagnetic mass oscillating above a base on
which magnets are placed at different positions, as illustrated in Fig.~\ref{fig:magpend-scheme}.
The interaction between the moving mass and the magnets produces trajectories that depend on
the initial conditions and on the configuration chosen for the base. This kind of system is
connected with the study of the magnetic pendulum, an example used to discuss chaotic
behaviour, sensitivity to initial conditions and the practical difficulty of
prediction~\citep{christian2020chaos}. Although the system obeys deterministic laws, small
differences in the initial position can lead to markedly different trajectories, a
characteristic feature of many chaotic
systems~\citep{strogatz2018nonlinear,baker1996chaotic}.

\begin{figure}[htbp]
  \centering
  \includegraphics[width=0.40\textwidth]{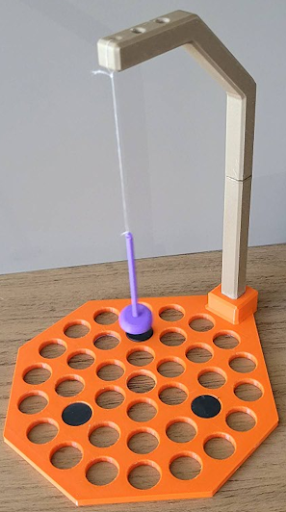}
  \caption{3D-printed magnetic pendulum, with a perforated base that allows the distribution
           of the magnets to be modified. Own elaboration.}
  \label{fig:magpend-3d}
\end{figure}

From the design point of view, 3D printing makes it possible to manufacture a stable base with
well-defined positions for inserting or removing magnets, as well as a support that fixes the
suspension point of the pendulum. This configuration makes it easy to modify the distribution
of the magnets and to explore how those changes affect the interaction with the suspended
mass. In this way, the printed device turns the setup into a controllable experimental
situation, in which the geometry of the base and the position of the magnets become variables
of study. Figure~\ref{fig:magpend-3d} shows a 3D-printed version of the magnetic pendulum.

In the classroom, the device can be used to compare the motion of the free pendulum with the
motion of the same system once the magnets are added. An initial task consists in asking
students to anticipate the expected trajectory, to sketch the possible path, and to state
which variables they would need to control in order to repeat the trial. Then, after
performing several repetitions with and without magnets on the base, similarities and
differences between trajectories can be discussed. This way of working makes it possible to
introduce sensitivity to initial conditions without starting from a formal definition of
chaos. Subsequently, the analysis can be extended through video recordings or motion-tracking
software, a resource already used in teaching sequences on deterministic
chaos~\citep{parlati2024design}.

\subsection{Woodpecker toy}

\begin{figure}[htbp]
  \centering
  \includegraphics[width=0.55\textwidth]{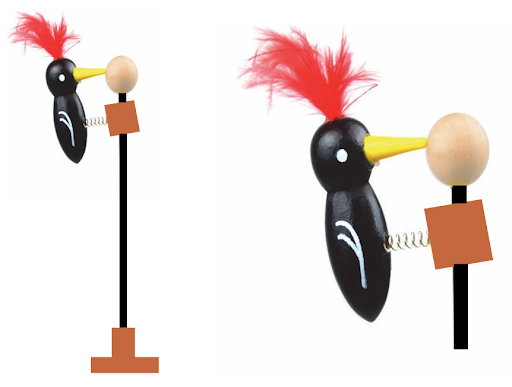}
  \caption{Representation of the woodpecker toy and detail of the body--spring--pole system.
           Reproduced from Ref.~\citep{rajasekar2023understanding}.}
  \label{fig:woodpecker-scheme}
\end{figure}

The woodpecker is a toy associated with German wooden toy making, known for producing an
unexpected descending motion from a very simple mechanism.
Figure~\ref{fig:woodpecker-scheme} shows a representation of the device and highlights the
main elements involved in its motion. In its most common version, the body of the bird is
attached by a spring to a small sleeve that surrounds a vertical pole. When the body is
displaced from its equilibrium position and released, the system starts to oscillate and
descends along the rod through the interaction between body, spring, sleeve and
pole~\citep{rajasekar2023understanding}. Despite its simple appearance, the system involves
conceptually rich mechanical phenomena. Depending on the initial conditions, the motion may be
damped until the toy comes to rest, or it may settle into a periodic descent, which makes it
possible to discuss how the response of the system depends on the amplitude of the vibration
and on the contact conditions~\citep{schlichting1987chaotic}. Furthermore, this toy has been
modelled as a nonlinear system with impacts and contact, well suited to analysing how changes
in the operating conditions modify the observed behaviour~\citep{leine2003nonlinear}.

\begin{figure}[htbp]
  \centering
  \includegraphics[width=0.40\textwidth]{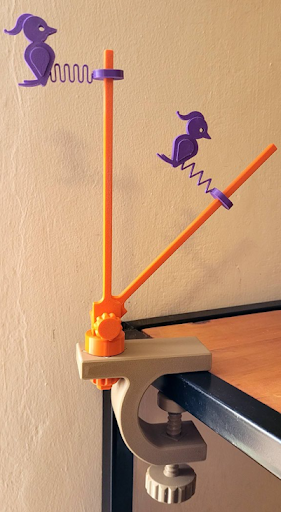}
  \caption{3D-printed woodpecker toys, with poles of different inclination and springs of
           different stiffness, used to compare the descent of the birds. Own elaboration.}
  \label{fig:woodpecker-3d}
\end{figure}

3D printing makes it possible to manufacture the body of the bird, to adjust its geometry and
to modify the position of the centre of mass or of the contact point with the pole. These
possibilities turn the toy into an adaptable experimental device, since small variations in
the design can modify its response during the descent. This way of working connects with the
use of physics toys as resources for exploring principles, models and operating
conditions~\citep{aravind2015inexpensive,rajasekar2023understanding}.
Figure~\ref{fig:woodpecker-3d} shows a 3D-printed version of the woodpecker, mounted on poles
with different inclinations and fitted with different springs.

Conceptually, the woodpecker makes it possible to analyse how certain parameters condition the
dynamics of the system. In particular, the stiffness of the spring and the inclination of the
pole influence the oscillation, the contact with the rod and the continuity of the descent.
Comparing different configurations therefore allows a discussion of how changes in the system
parameters can alter the speed, the regularity and the type of motion observed. This reading
recovers a central feature of the toy: from a simple mechanism, different dynamical responses
can emerge depending on the initial conditions and on the operating
parameters~\citep{leine2003nonlinear,schlichting1987chaotic,rajasekar2023understanding}.

One possible classroom activity consists in organising small competitions between devices,
modifying one parameter at a time. For example, birds with springs of different stiffness can
be compared on poles with the same inclination, or identical birds can be released on poles
with different inclinations. These comparisons make it possible to discuss which factors
influence the motion, what is being held constant, and how the speed, the regularity or the
continuity of the descent change. In this sense, the woodpecker connects the initial curiosity
with relevant questions about friction, impacts, oscillations and the dependence of the motion
on the parameters of the system~\citep{leine2003nonlinear,rajasekar2023understanding}.

\subsection{Double pendulum}

\begin{figure}[htbp]
  \centering
  \includegraphics[width=0.42\textwidth]{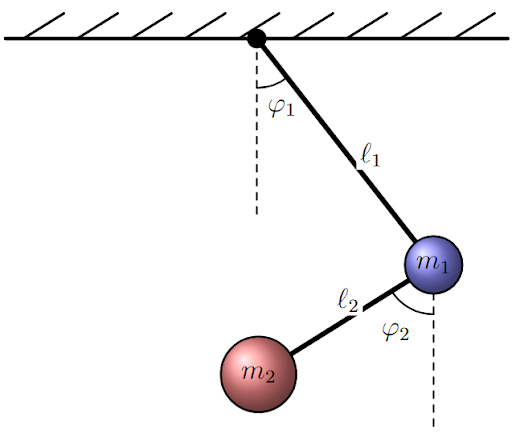}
  \caption{Idealised model of the double pendulum. Own elaboration.}
  \label{fig:dp-model}
\end{figure}

The double pendulum can be represented by an idealised model consisting of two point masses,
$m_1$ and $m_2$, joined by rigid, massless arms of lengths $\ell_1$ and $\ell_2$. The system
oscillates in a vertical plane and its configuration can be described by the angles
$\varphi_1$ and $\varphi_2$, as shown in Fig.~\ref{fig:dp-model}. Although this representation
simplifies the actual device, it allows a clear identification of the variables entering the
model.

The physical interest of the double pendulum does not lie in the complexity of its
construction, but in the richness of its motion. For small angular displacements and certain
initial conditions the system can display regular behaviour; for larger amplitudes, in
contrast, complex motions appear that are difficult to anticipate by direct observation. For
this reason the double pendulum is used as a paradigmatic example of a nonlinear dynamical
system, since it allows a discussion of the transition between regular and chaotic behaviour,
as well as of sensitivity to initial
conditions~\citep{strogatz2018nonlinear,baker1996chaotic,moon2004chaotic}.

\begin{figure}[htbp]
  \centering
  \includegraphics[width=0.40\textwidth]{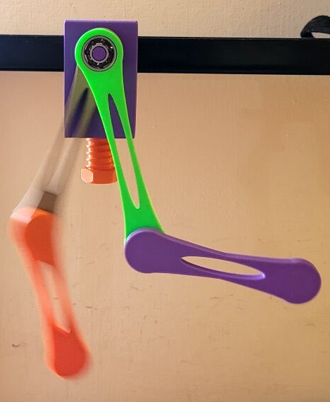}
  \caption{3D-printed double pendulums used to visualise the divergence of trajectories.
           Own elaboration.}
  \label{fig:dp-3d}
\end{figure}

3D printing makes it possible to manufacture lightweight arms, low-friction supports and
interchangeable parts. This makes it easy to modify lengths, masses or pivot points, so that
the same device can be used in activities at different levels of depth. As a first approach,
it can be used to observe qualitatively the transition from regular to more complex motion.
In a second stage, the motion can be recorded on video in order to analyse trajectories,
angular velocities, or comparisons between similar releases. This kind of record is
particularly relevant for teaching proposals that aim to bring students closer to
deterministic chaos through motion analysis~\citep{parlati2024design}.
Figure~\ref{fig:dp-3d} shows a setup with two 3D-printed double pendulums, used to visualise
the divergence between motions started from similar conditions.

In the classroom, the double pendulum can be used to discuss repetition, control and
prediction by means of a setup with two similar devices. One possibility consists in releasing
both pendulums from initial positions that are as close as possible and observing how their
motions evolve. Although at the beginning the trajectories may look similar, small differences
in the initial state of the system can be amplified with time and produce clearly distinct
motions. This visual divergence makes it possible to problematise the idea of exact
repeatability and to discuss the fact that knowing the laws governing a system is not always
enough to anticipate its evolution accurately over long time intervals, especially when the
system exhibits sensitivity to initial
conditions~\citep{strogatz2018nonlinear,baker1996chaotic}.

\section{Classroom guidelines}

Working with 3D-printed devices should not be reduced to presenting novel objects or striking
demonstrations. For these resources to acquire educational value, they must be integrated into
a plan that turns the device into a situation of inquiry, discussion and construction of
explanations. In this sense, the devices presented here can be articulated with active
learning methodologies, in which students formulate anticipations, observe phenomena, record
evidence, contrast ideas and elaborate interpretations. This approach is consistent with
research highlighting the role of active engagement in conceptual
understanding~\citep{hake1998interactive}, as well as with perspectives that take into account
students' initial ideas and the progressive construction of explanatory
models~\citep{redish2003teaching,duit2003conceptual}.

A particularly suitable way of organising these experiences is to structure the activities in
three stages: prediction, observation and explanation. In the prediction stage, students state
what they expect to happen and justify their anticipations before handling the device. This
step is relevant because it makes visible the initial assumptions with which they interpret
the system. In the observation stage, they carry out trials, compare repetitions and record
similarities or differences in the observed behaviour. In the explanation stage, they revisit
what was predicted and what was observed in order to construct a physical interpretation of
the phenomenon. This organisation follows the Predict--Observe--Explain approach, used to probe
students' understanding and to promote the revision of their initial
explanations~\citep{white2014probing}.

The focus of the activities should not merely be to observe that the motion is complex,
unexpected or hard to anticipate. The educational interest is enhanced when the tasks guide
the comparison between situations. In the magnetic pendulum, for example, comparing the free
motion with the motion under different magnet configurations allows a discussion of how the
response of the system changes when the interaction at the base is modified. In the woodpecker,
contrasting springs of different stiffness or poles with different inclinations allows an
analysis of how certain parameters condition the descent. In the double pendulum, the
comparison between two devices started from similar positions makes the progressive divergence
of their motions visible. In all cases, comparison shifts the attention from a superficial
description towards the identification of variables, parameters and operating conditions.

The wording of the tasks plays a central role in this orientation. It is not enough to ask
students to ``observe what happens''; it is necessary to pose questions that lead them to
anticipate, justify, record and revisit their own ideas. For example: which variable was
modified? what was held constant? what differences appear between two repetitions? does the
observed behaviour agree with what was predicted? which explanation best accounts for the
phenomenon? Questions of this kind help the device act as a mediator between the concrete
experience and the conceptual construction, preventing the activity from being reduced to a
demonstration.

Working in small teams is convenient because it favours peer discussion and the negotiation of
explanations. Before writing an answer, students can confront their predictions, agree on what
to observe and decide which aspects to record. Written records also fulfil an important
educational function: they preserve the initial anticipations and allow them to be compared
with what was observed. Activities may therefore include verbal descriptions, sketches of
trajectories, simple tables, configuration diagrams, or comparisons between trials. These
records are not intended to produce a single correct answer, but to gather evidence supporting
a physical explanation.

The role of the teacher is decisive in guiding the experience. Before the activity, the
teacher can help delimit the problem and specify which variable will be modified. During the
observation, the teacher can encourage repetitions, request comparisons and draw attention to
relevant aspects of the behaviour. After the experience, the teacher can organise a plenary
discussion that helps distinguish between description, explanation and model. In this closing
stage, questions such as ``what changed with respect to the initial prediction?'' or ``which
model best interprets what was observed?'' allow students to move from the visible phenomenon
towards a more elaborate understanding.

The depth of the work can be adjusted to the educational level and to the time available. In a
first approach, the devices allow a qualitative exploration centred on observation, comparison
and discussion. In longer sequences, video records, trajectory tracking, sensor measurements
or computational modelling can be incorporated. These extensions make it possible to move from
an initial phenomenological experience to a more systematic analysis of variables, parameters
and models. In particular, the use of video and tracking software can favour the study of
trajectories and an experimental approach to ideas related to deterministic
chaos~\citep{parlati2024design}.

An integrating closure is especially important when several devices are used within the same
sequence. Comparing the magnetic pendulum, the woodpecker and the double pendulum makes it
possible to recognise common features: dependence on parameters, sensitivity to small
variations, practical limits of prediction, and the relation between the configuration of the
system and the observed behaviour. In this way, 3D-printed devices can contribute to
introducing notions of dynamical systems and nonlinear physics from an experimental
perspective, without starting from advanced mathematical formalism, but without giving up the
progressive construction of physical models.

\section{Implementation possibilities, scope and limitations}

The 3D-printed devices presented in this work can be incorporated at different educational
levels, with varying degrees of conceptual and experimental depth. In secondary education,
they are suitable for introducing qualitative discussions about motion, forces, energy and
predictability. In pre-service teacher training, they offer an opportunity to revisit how
dynamics is taught, and to problematise the transition from a school physics centred on
idealised, linearised and highly predictive models towards a view that includes nonlinear
deterministic systems, sensitivity to initial conditions and practical limits of prediction.
In introductory university courses, they can be complemented with video records, trajectory
tracking or computational modelling, which allows notions of nonlinear physics to be
introduced from experimental work and to advance towards more formal treatments based on
graphical analysis, simulation and the discussion of
models~\citep{strogatz2018nonlinear,parlati2024design}.

The depth of the activities can be adjusted to the available time and to the teaching goals. As
a first approach, the devices can act as qualitative triggers for observing phenomena,
formulating predictions and comparing behaviours. In longer sequences, the work can be extended
through written records, video analysis, comparison of configurations and the elaboration of
models. This flexibility makes it possible to move from an initial exploration of the
phenomenon to a more systematic discussion of variables, parameters and operating conditions.

In project-based courses or workshops, the design, printing and improvement of the devices can
themselves become part of the learning process. Students can analyse which part should be
modified, which physical parameter they want to explore, or which operating condition needs to
be controlled. In this way, 3D printing appears not only as a manufacturing tool, but as an
opportunity to link design, experimentation and physical reasoning.

The main strength of this proposal is related to accessibility. Digital fabrication makes it
possible to build low-cost resources, replace damaged parts, adjust dimensions and share files
among teachers and institutions~\citep{coakley2016printing,mcnair2024toward}. This feature is
especially valuable when one wishes to work with phenomena for which school laboratories do not
usually have dedicated equipment. In addition, the possibility of modifying the design allows
the same device to be adapted to different teaching purposes, educational levels and class
schedules.

A further strength is conceptual. With these devices it is possible to introduce ideas
associated with dynamical systems, nonlinearity, parameter dependence, sensitivity to initial
conditions and limited predictability without starting from advanced mathematical treatment.
The concrete experience can act as a starting point for formulating questions, contrasting
predictions and constructing explanations. This mediation relies on the potential of physics
toys to promote conceptual discussion and on the organisation of activities around prediction,
observation and
explanation~\citep{aravind2015inexpensive,leine2003nonlinear,white2014probing,rajasekar2023understanding}.

The proposal also has limitations. One of them concerns access to 3D printers and the time
required to design, print and adjust the parts. Although the cost per unit can be low, the
process demands preliminary tests and a certain technical knowledge about slicing, print
orientation, tolerances and material strength. In addition, several devices require non-printed
components, such as strings, magnets or bearings, whose choice influences the final behaviour.

A second limitation is related to the reproducibility of the behaviour. In systems that are
sensitive to small variations, minor differences in the assembly can modify the observed
response. This variability should not be interpreted solely as a technical difficulty; it can
also become part of the phenomenon to be discussed, provided that the teacher defines in advance
which aspects are to be compared. In the magnetic pendulum, for instance, the position of the
magnets, the suspension height and the release mode all affect the trajectory. In the
woodpecker, the stiffness of the spring, the inclination of the pole and the mass distribution
condition the descent~\citep{leine2003nonlinear,christian2020chaos,rajasekar2023understanding}.

Finally, care must be taken with the interpretation constructed in the classroom. These devices
can arouse curiosity, but that curiosity needs to be guided by clear tasks. When the activity is
reduced to observing that the motion is strange or hard to anticipate, part of its educational
potential is lost. Tasks should therefore promote comparisons, identification of variables,
recording of repetitions and discussion of explanations. The role of the teacher is decisive in
linking the observed behaviour with physical concepts and in preventing the device from being
reduced to an eye-catching demonstration.

To facilitate the reproduction and adaptation of this proposal, the STL files, printing
parameters, bill of materials and assembly guidelines are available in an open
repository~[\textit{link}]. This material allows other teachers to print, modify and adapt the
devices to their own teaching contexts.

\section{Conclusions}

3D printing offers a concrete possibility for broadening the repertoire of experimental devices
available in physics teaching laboratories. The examples presented here show that it is possible
to design and build accessible, modifiable resources tailored to different teaching purposes.
Their main value lies not only in the fabrication of parts, but in the construction of classroom
situations in which students can observe, compare, record and explain physical phenomena. This
perspective is consistent with work that highlights the potential of digital fabrication for
producing open, adaptable and low-cost experimental
resources~\citep{coakley2016printing,mcnair2024toward}.

The magnetic pendulum, the woodpecker and the double pendulum make it possible to introduce ideas
associated with dynamical systems, parameter dependence, sensitivity to initial conditions and
limited predictability. These phenomena can be approached through a first qualitative
approximation, supported by manipulable devices, before moving towards more formal mathematical
or computational models.

The proposal underlines the importance of linking the construction of the device with the
planning of the teaching sequence. For a 3D-printed resource to work as a teaching resource in
physics, it must be embedded in activities with clear tasks, comparison of situations, recording
of observations and conceptual discussion. Within this framework, strategies such as POE, video
analysis or guided modelling can favour the transition from initial observation towards more
elaborate physical explanations~\citep{parlati2024design,white2014probing}.

As a future direction, it seems desirable to move towards open repositories of designs, assembly
guides and classroom guidelines that can be shared, reviewed and improved by teachers. In this
way, 3D printing can contribute to a culture of collaborative work in experimental physics
teaching, in which teachers take part in the creation and adaptation of devices suited to their
own contexts.

\section*{Data availability}

The STL files, printing parameters, bill of materials and assembly instructions for the three
devices are openly available at~[\textit{repository link}].

\bibliographystyle{unsrtnat}
\bibliography{refs}

\end{document}